# OPERATIONS OF THE TEVATRON ELECTRON LENSES*

X.L. Zhang#, V. Shiltsev, A. Valishev, G. Stancari, G. Kuznetsov, G. Saewert,
FNAL, Batavia, IL 60510, U.S.A.
V. Kamerdzhiev, FZ-Jülich, IKP, Germany

*Abstract*

The two Tevatron Electron Lenses (TEL1 and TEL2) are installed in Tevatron in 2001 and 2006 respectively. TEL1 is operated as the vital parts of the Tevatron for abort gap beam clearing, while TEL2 is functioning as the backup as well as the test device for beam-beam compensation, space charge compensator and soft beam collimator. Both of them are working exceptionally reliable after a few initial kinks being worked out. Their operations in Tevatron are summarized in this report.

## INTRODUCTION

The Electron Lenses have been installed in the Tevatron with the objective to compensate the beam-beam effects on antiproton beams which may limit the collider performance [1, 2]. The electron-beam current can be adjusted bunch-by-bunch to optimize the performance of all bunches in a multi-bunch collider by using fast high voltage modulator [6]. In addition, the electron transverse current profile (and thus the radial dependence of electromagnetic (EM) forces due to electron space-charge) can also be changed for different applications using different electron guns.

However for the present Tevatron operations, the antiproton beam lifetimes are dominated by luminosity consumptions [2], which leave the TEL a primary vital function as abort gap beam cleaner.

## TEVATRON ELECTRON LENSES

Both Tevatron Electron Lenses (TELs) direct their beam against the antiproton flow. Figure 1 shows the layout of the TEL2.

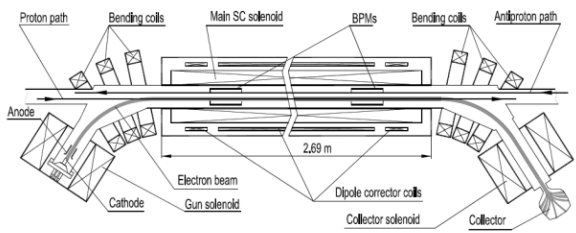

Figure 1: TEL2 layout.

Extensive studies have been carried out with electron beam profiles of flattop, smooth edge flattop (SEFT) and Gaussian in the Tevatron. The SEFT gun has been designed and built in order to generate much less nonlinearity than the flattop gun at the transit edges so that it causes much less proton loss when electron beam is not perfectly aligned with proton beam it acted on. The Gaussian gun was installed to study the nonlinear beam-beam compensation effects. Recently we have installed the hollow electron gun to study the electron beam collimations.

___________________________________________
*Work supported by the Fermi Research Alliance, under contract DE-AC02-76CH03000 with the U.S. Dept. of Energy.
#zhangxl@fnal.gov

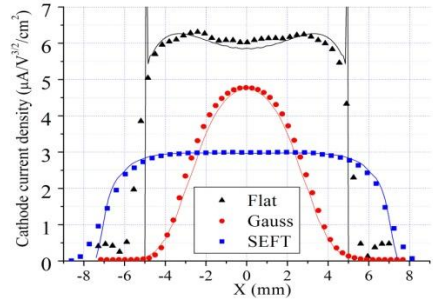

Figure 2: Three profiles of the electron current density at the electron gun cathode: black, flattop profile; red, Gaussian profile; blue, SEFT profile. Symbols represent the measured data and the solid lines are simulation results. All data are scaled to refer to an anode–cathode voltage of 10 kV.

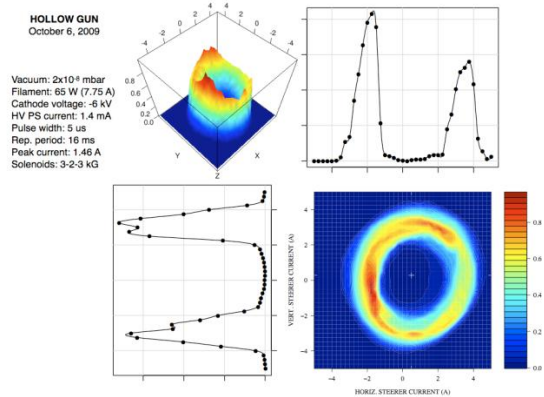

Figure 3: Current profiles of the hollow electron gun.

One example of the hollow electron beam profile was measured with a 0.2 mm pinhole and reconstructed as in Figure 3. To make this hollow cathode, a 0.6 inch spare SEFT gun cathode was bored out about 3/8 inch hole in the center. And about 2.5 A peak electron current achieved at 9 KV anode to cathode voltage.

## REMOVING UNCAPTURED BEAM

In Tevatron, uncaptured protons are generated by various mechanisms such as coalescing in the MI, intra-beam scattering (IBS), phase and amplitude noise of the RF voltage. The longitudinal instabilities or trips of the RF power amplifiers can contribute large spills of particles to the uncaptured beam. Uncaptured beam particles are outside of the RF buckets, and therefore, move longitudinally relative to the main bunches to fill the beam abort gap. If the number of particles in the uncaptured beam is too large and eventually lost due to energy ramp, beam abort or fallout, usually causing large background in physics detector, damage their components even lead to quenches of the superconducting (SC) magnets by the corresponding energy deposition.

To remove the uncaptured beam effectively and quickly, the TEL electron beam is timed to the abort gaps and placed transversely near beam orbit. Then the electron beam is turned on to excite dipole motion resonantly until they are lost. The flattop and SEFT electron beam have been already demonstrated effectively for this purpose in daily operation [7]. The effectiveness of the new Gaussian was also measured as show in Figure 4. The TEL was turned off during a store (average electron current is shown in black) at about 7 min. Accumulation of the uncaptured beam started immediately and can be measured by the Abort Gap Monitor (AGM) plotted in red. It grows for about 25 minutes reaching near saturation at intensity of about $3.8 \times 10^9$ protons. When the TEL2 turned back on, the accumulated protons were quickly cleared out. The similar cleaning effects were also confirmed with the newly installed hollow electron gun.

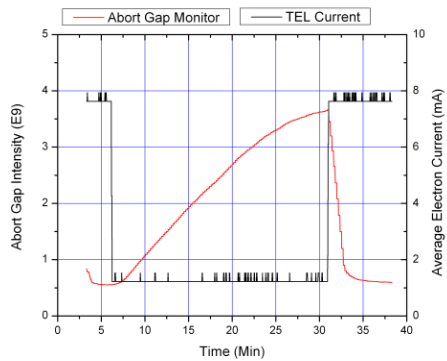

Figure 4: Uncaptured beam accumulation and removal by the TEL2. The black line represents the average electron current of the TEL; the red line is the uncaptured beam AGM.

## BEAM-BEAM COMPENSATION STUDIES

The experimental beam-beam compensation (BBC) studies [3] were carried out at the Tevatron for either dedicated machine time or parasitically during the High Energy Physics (HEP) store and mostly done with protons. The tune shift, beam lifetime and halo loss rate at both physics detectors are measured and analyzed.

### Tune Shifts

Figure 5 presents the vertical tune shift induced by the TEL2 electron current from the SEFT gun. There is an excellent agreement between the tune shift measured by the 1.7 GHz Schottky tune monitor and the theory. The dependence of the tune shift on the electron energy also agrees with the theoretical predictions

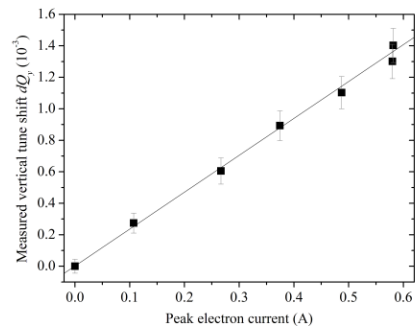

Figure 5: Vertical betatron tune shift of the 980 GeV proton bunch vs. the peak electron current in TEL2.

The studies have been also carried out for the Gaussian electron beam with antiproton beam at TEL2 recently [4]. In Figure 6 the vertical Schottky spectrum is plotted for one of the measurements. The observed tune shifts and enhancement of the tune spread are well expected.

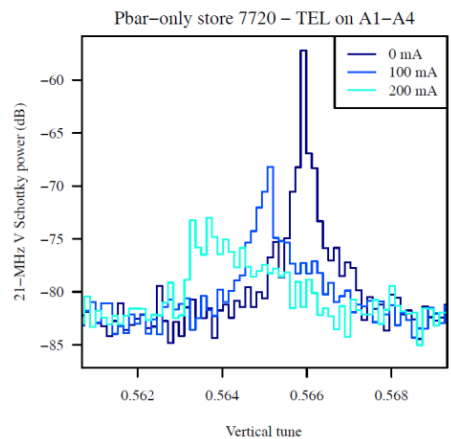

http://hb2010.web.psi.ch/authorinfo/prst.php

Figure 6: Antiproton beam vertical tune spread vs. Gaussian electron beam peak current in TEL2.

The studies also show that no deleterious effects of the Gaussian electron beam on the antiproton beams.

### Beam Lifetime

Usually, the proton lifetime, dominated by beam–beam effects, gradually improves with time in a HEP store due to the decreasing of the antiproton intensity and growing of antiproton emittance. In store #5119, effectiveness of the BBC was studied the by repeatedly turning on and off TEL2 on a single bunch P12 every half-an-hour for 16 h. The relative bunch intensity lifetime improvement $R$ is plotted in Figure 7.[5]

The first two data points correspond to $J_e = 0.6$ A, but subsequent points were taken with $J_e = 0.3$ A to observe the dependence of the compensation effect on the electron current. The change of the current resulted in a drop of the relative improvement from $R = 2.03$ to 1.4. A gradual decrease in the relative lifetime improvement is visible until after about 10 h, where the ratio reaches 1.0 (i.e. no gain in the lifetime). At this point, the beam–beam effects have become very small, providing little to compensate. Similar experiments in several other stores with initial luminosities ranging from $1.5\times10^{32}$ to $2.5\times10^{32}$ cm$^{-2}$s$^{-1}$ reproduced these results.

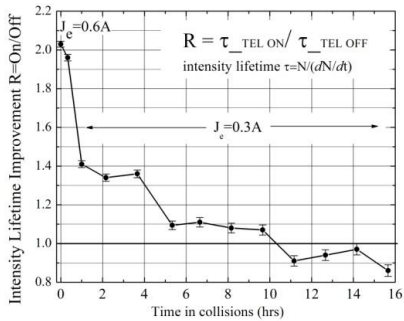

Figure 7: Relative improvement of the TEL2 induced proton bunch #12 lifetime vs. time (store #5119, Dec. 12, 2006, initial luminosity $L = 159 \times 10^{30}$ cm$^{-2}$ s$^{-1}$).

## COLLIMATION EFFECT

For particles interacting with the nonlinear force of the electron space charge field, they may be eventually lost due to their nonlinear dynamics. This phenomenon was observed experimentally with the flattop electron gun where the electron beam's sharp edges acted as a `gentle' collimator [7]. When proton beam passed through the electron beam, the outlying particles were slowly driven out of the bunch until they eventually hit the beam aperture. Encouraged by this observation, the hollow electron gun was developed and installed in the TEL2 to test the proposal of using hollow electron beam for the near beam collimation for LHC [10, 11].

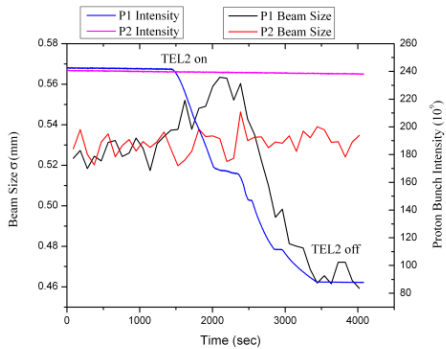

Figure 8. Scraping of a proton bunch due to interaction with the TEL2 electron beam (transverse hollow electron current distribution). The beam size was measured by synchrotron light monitor.

The very preliminary observations of the hollow electron beam collimation effects were carried out during the electron beam position scan for abort gap cleaning studies and were shown in Figure 8. The electron beam pulse was timed at proton bunch P1 while bunch P2 was used as the reference bunch. When the electron beam was turned on, the vertical beam size of P1 was increased a little at first. Then as the electron beam was scanned vertically cross the proton beam, the tail of the proton beam was scraped away. P1 intensity decayed fast as it was collimated by the hollow electron beam. The final equilibrium beam sized was not reached during the brief scan. The further detailed studies of the beam dynamics will be carried out soon.

## ELECTRON COLUMNS

The space charge effect is one of the main factors to limit intensity of proton beam in proposed high current proton storage rings. It could be compensated by sufficient number of "electron columns" devices which are capable of trapping electrons, generated from the ionization of residual gas by proton beam [9]. The strong longitudinal magnetic field of a solenoid is used to keep electrons from escaping from the transverse position they are born at and suppress the e-p instability while at the same time weak enough to allow ions escape and not affect the process of charge compensation. Two ring electrodes at both ends of the solenoid supply electric field to trap the electrons longitudinally.

The preliminary studies using the Tevatron Electron Lens configured to work as "electron column" had shown significant accumulation of electrons inside an electrostatic trap in 3T longitudinal magnetic field with intentionally increased vacuum pressure. These negatively charged electrons moved vertical tune of 150 GeV proton beam upward by as much as +0.005.

And the significant vacuum instability was observed accompanied by the proton beam instability. Further theoretical and bench studies are needed to understand the dynamic processes inside the ionized and magnetized "electron column".

## SUMMARY

The effects of BBC were successful demonstrated in the Tevatron. And various useful applications have been also studied before [8]. The abort gap cleaning ability is also proven to be vital to the daily operations. Now the hollow electron gun was installed to examine the collimation effect in details.

## REFERENCES

[1] V.Shiltsev *et al.*, Phys. Rev. ST Accel. Beams 8, 101001 (2005).
[2] V.Shiltsev, *et al.*, Phys. Rev. ST Accel. Beams 11, 103501 (2008).
[3] V.Shiltsev, *et al.*, New J. Phys. 10 (2008) 043042.
[4] A. Valishev, *et al.*, IPAC10, pp.2084-2086, 2010.
[5] V.Shiltsev, *et al.*, Phys. Rev. Lett. 99 244801 (2007).
[6] G.W. Saewert, *et al.*, PAC09, TU6RFP079, 2009.
[7] Xiao-long Zhang, *et al.*, Phys. Rev. ST Accel. Beams 11, 051002 (2008).
[8] X. Zhang, et al., Proceedings of the 2003 Particle Acc. Conf. pp 1778-1780, Portland, USA. 2003.
[9] V.Shiltsev, *et al.*, PAC09, TH5PFP020, 2009.
[10] J. C. Smith, *et al.*, PAC09, WE6RFP031, 2009.
[11] G. Stancari, *et al*., FERMILAB-CONF-10-196-APC